\documentclass[useAMS,usenatbib]{mn2e}

\usepackage{graphicx}

\newcommand       \be           {\begin{equation}}
\newcommand       \ee           {\end{equation}}
\newcommand       \bea          {\begin{eqnarray}}
\newcommand       \eea          {\end{eqnarray}}
\newcommand       \apj          {ApJ}
\newcommand       \apjl         {ApJL}
\newcommand       \aap          {A\&A}
\newcommand       \nat          {Nature}
\newcommand       \mnras        {MNRAS}
\def\simlt{\mathrel{\hbox{\rlap{\hbox{\lower4pt\hbox{$\sim$}}}\hbox{$<$}}}}
\def\simgt{\mathrel{\hbox{\rlap{\hbox{\lower4pt\hbox{$\sim$}}}\hbox{$>$}}}}

\title[Extended Emission from Proto-Magnetar Spin-Down]{Short Duration Gamma-Ray Bursts with Extended Emission from Proto-Magnetar Spin-Down} \author[B.D. Metzger, E. Quataert, T.A. Thompson]{B.~D. Metzger$^{1,2}$\thanks{E-mail:
bmetzger@astro.berkeley.edu}, E. Quataert$^{1}$, T.~A. Thompson$^{3}$\\
$^{1}$Astronomy Department and Theoretical Astrophysics Center,
University of California, Berkeley, 601 Campbell Hall, Berkeley CA,
94720\\ $^{2}$Department of Physics, University of California,
Berkeley, Le Conte Hall, Berkeley, CA 94720 \\ $^{3}$Department of Astronomy and Center for Cosmology $\&$ Astro-Particle Physics, The Ohio State University, 140 W. 18th Ave., Columbus, OH, 43210\\}
\begin{document}
\date{Accepted . Received ; in original form }
\pagerange{\pageref{firstpage}--\pageref{lastpage}} \pubyear{????}
\maketitle
\label{firstpage}

\begin{abstract}
Evidence is growing for a class of gamma-ray bursts (GRBs) characterized by an initial $\sim 0.1-1$ s spike of hard radiation followed, after a $\sim 3-10$ s lull in emission, by a softer period of extended emission lasting $\sim 10-100$ s.  In a few well-studied cases, these ``short GRBs with extended emission'' show no evidence for a bright associated supernova (SN).  We propose that these events are produced by the formation and early evolution of a highly magnetized, rapidly rotating neutron star (a ``proto-magnetar'') which is formed from the accretion-induced collapse (AIC) of a white dwarf (WD), the merger and collapse of a WD-WD binary, or, perhaps, the merger of a double neutron star binary.  The initial emission spike is powered by accretion onto the proto-magnetar from a small disk that is formed during the AIC or merger event.  The extended emission is produced by a relativistic wind that extracts the rotational energy of the proto-magnetar on a timescale $\sim 10-100$ s.  The $\sim 10$ s delay between the prompt and extended emission is the time required for the newly-formed proto-magnetar to cool sufficiently that the neutrino-heated wind from its surface becomes ultra-relativistic.  Because a proto-magnetar ejects little or no $^{56}$Ni ($< 10^{-3}M_{\sun}$), these events should not produce a bright SN-like transient.  We model the extended emission from GRB060614 using spin-down calculations of a cooling proto-magnetar, finding reasonable agreement with observations for a magnetar with an initial rotation period of $\sim 1$ ms and a surface dipole field of $\sim 3\times 10^{15}$ G.  If GRBs are indeed produced by AIC or WD-WD mergers, they should occur within a mixture of both early and late-type galaxies and should not produce strong gravitational wave emission.  An additional consequence of our model is the existence of X-ray flashes unaccompanied by a bright SN and not associated with massive star formation.
\end{abstract}


\vspace{-0.7cm}
\section{Introduction}
\label{sec:int}
\voffset=-2cm
\vspace{-0.2cm}
\emph{Swift} and HETE-2 have demonstrated that long and short duration gamma-ray bursts (GRBs) originate from distinct stellar progenitors.  Long GRBs track ongoing star formation and result from the death of massive stars (Stanek et al.~2003).  On the other hand, short GRBs have now been localized to both early and late-type host galaxies of moderate redshift (Barthelmy et al.~2005; Berger et al.~2005; Bloom et al.~2006), indicating a more evolved progenitor population.

Even among the classes of ``long'' and ``short'' GRBs, however, diversity is emerging.  One example is GRB060505 detected by the Burst Alert Telescope (BAT) on-board \emph{Swift}.  This long duration burst showed no evidence for a bright supernova (SN) despite residing inside a star-forming region (Ofek et al.~2007; Fynbo et al.~2007), suggesting that it may result from a ``failed-SN'' collapsar (Woosley 1993).  Another particularly striking example is GRB060614, which, although officially classified as a long-duration burst based solely on its $102$ s T$_{90}$ duration, more closely resembles a standard short GRB in other ways.  This \emph{Swift} burst showed no energy-dependent time lag (Gehrels et al.~2006) and, like GRB060615, no SN was detected down to an optical brightness $\sim 100$ times fainter than SN1998bw (Gal-Yam et al.~2007).  Additional clues to the nature of GRB060614 are revealed through the evolution of its prompt emission.  The BAT lightcurve begins with an initial spike of hard emission (lasting $\sim 5$ s) which is followed, after a lull in emission (lasting $\sim 5$ s), by a softer `hump' of extended emission (lasting $\sim 100$ s).  This is followed by a remarkably standard X-ray, optical, and ultra-violet afterglow (Mangano et al.~2007).

The hybrid long/short properties of GRB060614 led Gal-Yam et al.~(2006) and Gehrels et al.~(2006) to propose that it forms the prototype for a new class of GRBs, which we call ``short GRBs with extended emission'' (or SGRBEEs) (see, however, Fynbo et al.~2006).  Roughly a quarter of the short bursts detected by \emph{Swift} (including GRBs 050709 and 050724) show evidence for high energy extended emission (EE) distinct from the standard afterglow and late-time flares; this suggests that SGRBEEs may actually be fairly common.  Indeed, Norris $\&$ Bonnell (2006; NB06) found a handful of short GRBs in the BATSE catalog qualitatively similar to GRB060614.  Although NB06's sample represents only $\sim 1\%$ of the BATSE short bursts, a soft tail of EE would generally not have been detectable.  NB06 also find that the dynamic range in the ratio of prompt to extended flux (and fluence) of SGRBEEs appears to be very large, $\sim 10^{4}$.  This large burst-to-burst variation in the relative energy released during the prompt and extended phases suggests that these components are physically decoupled.

One explanation for EE from an otherwise short GRB is the interaction of the relativistic outflow with the circumburst environment.  However, when the EE is sufficiently bright to be accurately sampled, its time evolution is highly variable and cannot be smoothly extrapolated back from the onset of the X-ray afterglow (Nakar 2007).  A multi-peaked lightcurve is also difficult to produce from the shock heating of a binary companion (MacFadyen et al.~2005).  Based on its similarity to prompt emission, the EE in SGRBEEs most likely results from late-time central engine activity.

A popular model for the central engine of short GRBs is accretion onto a black hole (BH) formed from a compact object merger (COM) (Paczynski 1986).  SGRBEEs pose a challenge to COM scenarios because their long durations and two-component nature are difficult to produce in models powered purely by accretion.  The accretion timescale of the compact disk produced from a merger event, although comparable to the duration of the initial spike, cannot explain the long duration of the EE, especially in cases when the latter produces the bulk of the observed fluence.  For BH-NS mergers, the fall-back of matter ejected into highly eccentric orbits during the NS's tidal disruption may be sufficient to power the EE (Rosswog 2007), but whether the regular delay between the prompt and EE phases, and the large variation in the flux of each, can be reproduced remains to be determined (see, however, Faber et al.~2006).

The NS kicks required to produce a compact binary and the potentially long delay until merger imply that a significant fraction of COMs should have large offsets from their host galaxies.  Although the offset distribution of short GRBs as a whole appears marginally consistent with current COM population synthesis models (Belczynski et al.~2006), well-localized SGRBEEs thus far appear exclusively inside or near the starlight of their host galaxies; indeed, their average offset from host center is only $\sim$ 2.5 kpc (Troja et al.~2007).  The high incidence of SGRBEEs with detected optical afterglows ($\sim 90 \%$) also suggests that these events reside inside the disk of their host galaxies (Troja et al.~2007).

In this Letter, we propose that SGRBEEs are produced by the formation and early evolution of a strongly magnetized, rapidly rotating neutron star (a ``proto-magnetar'') which is formed from the accretion-induced collapse (AIC) of a white dwarf (WD) (Nomoto $\&$ Kondo 1991), the merger and collapse of a WD-WD binary (King et al.~2001), or, perhaps, the merger of a double neutron (NS) star binary (Gao $\&$ Fan 2006).  The initial spike of emission is powered by accretion onto the proto-magnetar from a small disk formed during the AIC or merger (\S \ref{sec:accretion}).  The EE is produced by a relativistic wind that extracts the rotational energy of the magnetar on a timescale $\sim 10-100$ s (\S \ref{sec:spindown}), a picture similar to that originally proposed by Usov (1992).  The $\sim 3-10$ s delay between the prompt and EE is the time required for the proto-magnetar to cool sufficiently that the neutrino-heated wind from its surface becomes ultra-relativistic.  In \S \ref{sec:discussion} we summarize our results and the predictions of our model.  For concreteness we focus our discussion on AIC and WD-WD merger channels of isolated magnetar birth.
\vspace{-1.0cm}
\section{Accretion Phase}
\label{sec:accretion}
\vspace{-0.2cm}
In either AIC (Nomoto $\&$ Kondo 1991) or a WD-WD merger (e.g., Yoon et al.~2007), the WD (or merged WD binary) will be rapidly rotating prior to collapse and must eject a sizable fraction of its mass into a disk during the collapse in order to conserve angular momentum.  Indeed, the 2D MHD AIC calculations performed by Dessart et al.~(2007; D07) show that a quasi-Keplerian $\sim 0.1-0.5M_{\sun}$ accretion disk forms around the newly-formed proto-neutron star (PNS), extending from the PNS surface at $R_{\rm NS} \sim 30$ km to large radii (with a half-mass radius of a few $R_{\rm NS}$).

We propose that the prompt emission in SGRBEEs is powered by the accretion of this disk onto the PNS.  This scenario is not unlike most other COM models with the important exception that the accreting object is a NS rather than a BH.  The characteristic timescale for accretion is the viscous timescale $t_{\rm visc} = R^{2}/\alpha\Omega_{K}H^{2}$, given by
\begin{eqnarray}
t_{\rm visc} \approx 1{\rm\,s\,} \left(\frac{M}{M_{\sun}}\right)^{-1/2}\left(\frac{0.1}{\alpha}\right)\left(\frac{R_{0}}{4R_{\rm NS}}\right)^{3/2}\left(\frac{H/R_{0}}{0.2}\right)^{-2},
\label{tvisc}
\end{eqnarray}
where $\alpha$ is the viscosity parameter and $M$ is the PNS mass; $R_{0}$, $H$, and $\Omega_{\rm K} \equiv (GM/R_{0}^{3})^{1/2}$ are the disk's radius, scale height, and Keplerian rotation rate, respectively.  For $H \approx 0.2R_{0}$, as expected for a neutrino-cooled disk accreting at $\dot{M} \sim 0.1-1\,M_{\sun}$ s$^{-1}$ (Chen $\&$ Beloborodov 2007), $t_{\rm visc} \sim 0.1-1$ s, comparable to the duration of the prompt spike.  

At early times, the accretion ram pressure $P_{\rm R} \simeq \rho v_{\rm ff}^{2}/2 \approx \dot{M}v_{\rm ff}/8\pi R^{2}$ (where $v_{\rm ff} = (GM/R)^{1/2}$) exceeds the magnetic pressure $P_{\rm M} = B^{2}/8\pi$ at the PNS surface:
\be \frac{P_{\rm R}}{P_{\rm M}} \approx 10\left(\frac{\dot{M}}{0.1 M_{\sun}{\rm\,s^{-1}}}\right)\left(\frac{M}{M_{\sun}}\right)^{1/2}\left(\frac{\phi_{B}}{10^{27}{\rm G\,cm^{2}}}\right)^{-2}\left(\frac{R_{\rm NS}}{30{\rm\,km}}\right)^{3/2}.
\label{pressures}
\ee
Here $\phi_{B}=10^{27}(B(R_{\rm NS})/10^{15}{\rm G})(R_{\rm NS}/10{\rm\,km})^{2}$ G cm$^{2}$ is the NS's conserved magnetic flux.  Thus, although we postulate that the NS possesses a surface field strength $\sim 10^{15}$ G once contracting to its final radius $R_{\rm NS} \sim 10$ km, the field should not significantly alter the early-time dynamics of the accretion-powered phase (Ghosh $\&$ Lamb 1978).

The total energy released when a $\sim 0.1-0.5M_{\sun}$ disk accretes onto a NS ($\sim 10^{52}-10^{53}$ ergs) is more than sufficient to explain the isotropic $\gamma$-ray energy of the prompt spike of GRB060614 ($\simeq 1.8\times 10^{50}$ ergs).  However, as we discuss further in \S \ref{sec:spindown}, a major obstacle to driving a relativistic wind from the vicinity of a newly-formed PNS is the baryon-rich wind from the hot PNS's surface.  Because the mass loss from a rapidly rotating, highly magnetized PNS is augmented by centrifugal effects (Thompson et al.~2004; hereafter TCQ04), D07 argue that an early-time relativistic outflow from the PNS is unlikely.  Although we agree with D07's conclusion for moderately low latitudes, the centrifugal enhancement of mass loss along the rotation axis is negligible; hence, the total mass loss per solid angle near the rotation axis is approximately given by its non-rotating value of $M_{\Omega} \sim 10^{-5} M_{\sun}$ str$^{-1}$ (Thompson et al.~2001; hereafter T01).  Thus, if the energy deposited per solid angle above the pole exceeds $E_{\Omega} \sim 10^{51}$ ergs str$^{-1}$ the Lorentz factor of the outflow may reach $\Gamma \sim E_{\Omega}/M_{\Omega}c^{2} \sim 100$, sufficient to overcome compactness constraints which can be placed on short GRBs (Nakar 2007).

One possibility for effectively baryon-free energy deposition is neutrino-antineutrino annihilation along the rotation axis.  For instance, Setiawan et al.~(2006) find that $\sim 2\times 10^{50}$ ergs is released by annihilations from a $\sim 0.1 M_{\sun}$ disk accreting at $\dot{M} \sim 0.3 M_{\sun}$ s$^{-1}$, as would be expected following AIC.  An MHD jet is perhaps a more promising source of the relativistic material that produces the prompt emission.  Although jets from NS X-ray binaries are in general less powerful than their BH counterparts (Migliari $\&$ Fender 2006), the NS X-ray binary Circinus X-1 produces one of the most relativistic microquasar jets known (Fender et al.~2004).
\vspace{-0.9cm}
\section{Spin-Down Phase}
\label{sec:spindown}
\vspace{-0.2cm}
Whether produced by the core collapse of a massive star or the AIC of a WD, a PNS must radiate its gravitational binding energy via optically-thick neutrino emission during the first $t_{\rm KH} \sim 40$ s of its life (Burrows $\&$ Lattimer 1987).  A small fraction of this neutrino flux is reabsorbed by baryons in the PNS's atmosphere, driving a wind from its surface.  In the presence of a sufficiently strong magnetic field and rapid rotation, magnetic stresses tap into the PNS's rotational energy, enhancing the energy-loss in the wind (TCQ04).  The proto-magnetar is unlikely to have a significant effect on the accretion-powered phase for $t \simlt t_{\rm visc} \sim 1$ s (see eq.~[\ref{pressures}]).  On somewhat longer timescales, however, the disk mass and accretion rate will decrease, and the PNS will be spun up through accretion and by its Kelvin-Helmholtz contraction.  Thus, as $\dot{M}$ decreases from its peak value, the disk will be cleared away and the proto-magnetar wind will expand relatively freely into space.

In a previous work, we calculated the properties of PNS winds with magnetic fields and rotation during the first $t_{\rm KH} \sim 40$ s after core bounce (Metzger et al.~2007; M07a).  The importance of the magnetic field in accelerating a wind is quantified by the magnetization $\sigma \equiv B^{2}/4\pi\rho c^{2}$ evaluated at the light cylinder radius $R_{\rm L} \equiv c/\Omega$, where $\Omega$ is the PNS's rotation rate.  If the magnetic energy is fully converted into the kinetic energy of bulk motion, either directly or through thermalization and subsequent thermal or magnetic pressure-driven expansion (e.g., Drenkhahn $\&$ Spruit 2002), then at sufficiently large radii $\Gamma \sim \sigma(R_{\rm L}) \equiv \sigma_{\rm LC}$ (for $\sigma_{\rm LC} >1$).  Hence, for a PNS to produce an ultra-relativistic outflow, $\sigma_{\rm LC}$ must be $\gg 1$.  

Figure \ref{fig:wind} shows our calculation of $\sigma_{\rm LC}$ and the wind energy loss rate $\dot{E}$ as a function of time $t$ after core bounce for a PNS with initial rotation period $P_{0} = 1$ ms for three surface dipole field strengths: $B_{0} = 10^{15}$ G, $3\times 10^{15}$ G, and $10^{16}$ G.  In all models, the wind is nonrelativistic ($\sigma_{\rm LC} < 1$) for $t \simlt 1-10$ s because at early times the PNS is hot and its already substantial neutrino-driven mass-loss rate is enhanced by centrifugal slinging.  However, as the PNS cools and spins down, Figure \ref{fig:wind} shows that $\sigma_{\rm LC}$ rises rapidly, exceeding $\sim 10$ by $t \sim 3-10$ s.  Because an ultra-relativistic outflow is necessary to produce nonthermal GRB emission, the $\sim 3-10$ s timescale required for the wind to reach large $\sigma_{\rm LC}$ corresponds to the delay between the accretion-powered prompt spike and the spin-down-powered EE in our model.
\vspace{-0.9cm}
\subsection{Extended Emission Light Curve Model}
\label{sec:lightcurve}
\begin{figure}
\resizebox{\hsize}{!}{\includegraphics[bb = 0 0 500 360]{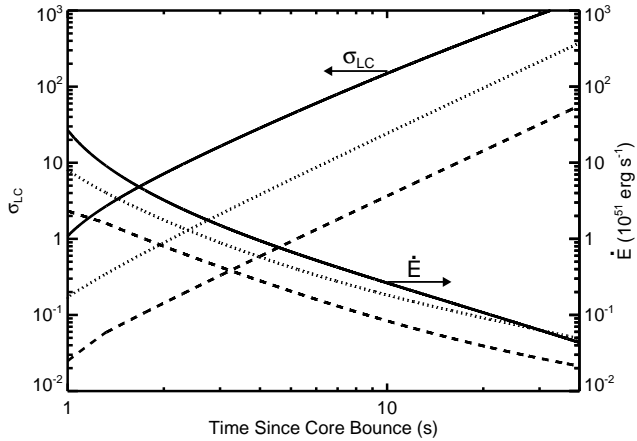}}
\caption{Magnetization at the light cylinder $\sigma_{\rm LC}$ (\emph{Left Axis}) and energy loss rate $\dot{E}$ (\emph{Right Axis}) as a function of time since core bounce for a proto-magnetar with initial rotation period $P_{0} = 1$ ms and three surface dipole magnetic field strengths: $B_{0} = 10^{15}$ G (\emph{dashed line}), $3\times 10^{15}$ G (\emph{dotted line}), and $10^{16}$ G (\emph{solid line}).}
\label{fig:wind}
\end{figure}

\vspace{-0.2cm}
In an attempt to directly connect central engine physics to observed GRB properties, we have modeled the emission produced by a spinning-down proto-magnetar using the wind evolution calculations (Fig.~\ref{fig:wind}) and an internal shock emission model.  By invoking strong shocks, we assume that the Poynting-flux of the wind is efficiently converted into kinetic energy somewhere between the light cylinder radius ($\sim 10^{7}$ cm) and the internal shock radii ($\sim 10^{13}-10^{15}$ cm).  This assumption is motivated by observations of the Crab Nebula and other pulsar wind nebulae, where detailed modeling requires efficient conversion of magnetic to kinetic energy (Kennel $\&$ Coroniti 1984).  Although we do not exclude the possibility that the wind remains Poynting-flux dominated at large radii, we leave the analogous light curve calculation in a magnetic dissipation model to future work. 

\begin{figure}
\resizebox{\hsize}{!}{\includegraphics[bb = 0 0 500 360]{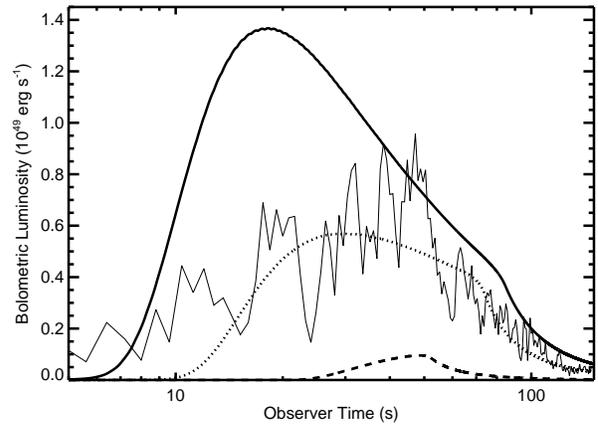}}
\caption{Luminosity of internal shock emission from the proto-magnetar winds in Figure \ref{fig:wind}; electron acceleration efficiency $\epsilon_{e} = 0.5$ is assumed.  Note the lack of emission at early times because the outflow is non-relativistic.  The gradual onset of the emission once $\sigma_{\rm LC} > 10$ is due to the large Thomson optical depth, which decreases as the outflow expands.  The late-time decline in emission is the onset of curvature emission from the last shock, produced by the shell released at $t_{\rm KH} = 40$ s.  The late-time BAT light curve from GRB060614, shown with a light solid line and scaled to the physical isotropic luminosity, is reproduced in a time-averaged sense by the $B_{0} = 3\times 10^{15}$ G model.}
\label{fig:lc}
\end{figure}

In the internal shock model, a GRB's emission is powered by ``dissipation'' (i.e., electron acceleration and radiation) of the relative kinetic energy between distinct components of a relativistic wind.  Since we do not have a quantitative model for the (potentially stochastic) processes that set the short timescale variability in a proto-magnetar wind, we discretize the relativistic outflow into $N$ shells released at constant intervals $dt$ in time.  A shell released at time $t$ is given Lorentz factor $\Gamma = \sigma_{\rm LC}(t)$, energy $E =\dot{E}(t)dt$, and mass $M = E/\Gamma c^{2}$, where $\sigma_{\rm LC}(t)$ and $\dot{E}(t)$ are taken from Figure \ref{fig:wind}.  The ``shell averaged'' light curve that we present in Figure \ref{fig:lc} is taken in the limit that $N \rightarrow \infty$ and $dt \rightarrow 0$ and is insensitive to the shell discretization prescription adopted.  We begin releasing shells when $\sigma_{\rm LC} > 10$ because the 2D MHD simulations of Bucciantini et al.~(2006) show that above $\sigma_{\rm LC} \sim 10$ the outflow transitions from being hoop-stress collimated along the rotation axis to expanding more isotropically; hence, we do not expect the material ejected at $\sigma_{\rm LC} \simgt 10$ to interact with much of the material ejected when $\sigma_{\rm LC} \simlt  10$.  We stop releasing shells when neutrino heating effectively ceases at $t = t_{\rm KH}$ because we do not have a reliable model for $\sigma_{\rm LC}(t)$ after this point.  This is a reasonable approximation if $\dot{E}$ or the dissipation efficiency drops significantly once the outflow transitions from a modest-$\sigma$ wind to a very high-$\sigma$, pulsar-like wind at $t=t_{\rm KH}$ (as suggested by, e.g., Thompson 1994 and TCQ04).

Upon release, each shell propagates forward in radius with constant velocity until it collides with another shell.  From the properties of the collision, we calculate (1) the ``thermal'' energy released by dissipation of the shells' relative kinetic energy, (2) the observed spike of radiation (using the technique summarized in \S 2 of Genet et al.~2007), and (3) the final mass and energy of the composite shell, which then continues to propagate forward.  We assume that a fraction $\epsilon_{e}$ of the energy released by each collision goes into relativistic electrons, which radiate their energy through synchrotron emission.  Efficient synchrotron cooling is justified if even a modest fraction of the magnetic flux at the light cylinder is preserved to large radii.  Thomson scattering of the nonthermal radiation is taken into account, but photospheric emission is not calculated.

Figure \ref{fig:lc} shows our calculation of the EE light curve for the wind solutions given in Figure \ref{fig:wind}.  We find that the efficiency for converting the relative kinetic energy of the outflow to thermalized energy is $\sim 10-20\%$.  Provided that $\epsilon_{e} \simgt 0.1$, these efficiencies are consistent with those typically inferred for short GRBs (e.g., Nakar 2007).  Proto-magnetar winds possess a significant reservoir of  ``free energy'' and achieve high efficiency because $\Gamma(t)$ increases monotonically, allowing faster material ejected at later times to catch up with the slower material ejected earlier.  

To first order, our simplified model produces light curves similar to the EE observed from SGRBEEs.  The peak flux is larger  for more rapidly rotating, strongly magnetized PNSs and the time to peak flux is smaller.  In Figure \ref{fig:lc} we also show the late-time BAT light curve from GRB060614 (from Butler $\&$ Kocevski 2007) for comparison with our models.  We find reasonable agreement between the data and the model with $B_{0} = 3\times 10^{15}$ G, suggesting that the progenitor of GRB060614 possesses a surface field strength somewhat larger than those of Galactic magnetars.  If synchrotron internal shock emission is indeed the correct model for the radiation from a proto-magnetar wind, the softening of the EE can also be qualitatively understood.  Due to the monotonic rise of $\Gamma(t)$, the Lorentz factor of the aggregate shell increases with time; however, the field strength in the wind $B \sim B(R_{\rm L})(r/R_{\rm L})^{-1}$ declines as the internal shock radius increases.  In our model, these effects combine to decrease the synchrotron peak energy $E_{\rm peak} \propto \Gamma B$ by a factor of $\sim 10$ during the period of observable emission.  This predicted degree of spectral softening is stronger than the factor of $\sim 2$ decrease in $E_{\rm peak}$ inferred for GRB060614 by Zhang et al.~(2007); indeed, the observed constancy of $E_{\rm peak}$ is a problem generic to most internal shock models.  
\vspace{-0.8cm}
\section{Discussion}
\label{sec:discussion}
Short GRBs with extended emission challenge the paradigm that short GRBs result exclusively from COMs.  The central engine in these systems may instead be a newly-formed magnetar.  The timeline of our model is summarized as follows:
\begin{itemize}
\item{AIC or WD-WD merger produces a proto-magnetar and a disk of mass $\sim 0.1M_{\sun}$ ($t\sim t_{\rm dyn} \sim 100$ ms)}
\item{Disk accretes onto the proto-magnetar, generating the prompt emission spike ($t\sim t_{\rm visc} \sim 0.1-1$ s; see eq.~[\ref{tvisc}])}
\item{ Free proto-magnetar wind transitions from non-relativistic to ultra-relativistic ($t \sim 3-10$ s; see Fig.~\ref{fig:wind}) }
\item{Proto-magnetar spins down, generating X-ray emission on observed longer timescale ($t \sim 10-100$ s; see Fig.~\ref{fig:lc})}
\end{itemize}
A model similar to the one described here was proposed by Gao $\&$ Fan (2006); in their model, late-time flares from short GRBs are powered by dipole spin-down of a super-massive, transiently-stable magnetar formed by a NS-NS merger.  However, current evidence suggests that SGRBEEs form a distinct population with only modest offsets from their host galaxies (Troja et al.~2007).  If transiently-stable magnetars from NS-NS mergers indeed produce most SGRBEEs, an equal number would be expected with large offsets.  

A more promising channel of isolated magnetar birth may be the AIC of a WD or the merger and collapse of a WD-WD binary.  The rate of these events is difficult to constrain directly because the Ni mass synthesized in a PNS wind is less than $\sim 10^{-3}M_{\sun}$ (Metzger et al.~2007b; M07b) and is therefore unlikely to produce a bright optical transient.  There is, however, indirect evidence that isolated magnetar birth occurs in nature.  The rapidly rotating, highly magnetic WD RE J0317-853 has a mass $M=1.35 M_{\sun}$ and was likely produced from a WD-WD merger; if RE J0317-853's progenitor binary had been slightly more massive, it would probably have collapsed to a rapidly rotating magnetar (King et al.~2001).  Isolated NS birth via AIC is also one of the only Galactic $r$-process sites consistent with current observations of elemental abundances in metal-poor halo stars (Qian $\&$ Wasserburg 2007).  Although unmagnetized PNS winds fail to produce successful $r$-process (T01), proto-magnetar winds may be sufficiently neutron-rich to produce $\sim 0.1M_{\sun}$ in $r$-process elements (D07; M07b).  For AIC or WD-WD mergers to produce the entire Galactic $r$-process yield requires a rate $\sim 10^{-5}-10^{-6}$ yr$^{-1}$, comparable to the observed local short GRB rate (Nakar 2007).  Finally, Levan et al.~(2006) argue that the correlation found by Tanvir et al.~(2005) between a subset of short GRBs and local large-scale structure is evidence for a channel of isolated magnetar birth, if these bursts are produced by SGR-like flares.

A theory for SGRBEEs must explain the large burst-to-burst variation in the ratio of the flux/fluence of the prompt and EE components (NB06).  The angular momentum of AIC and WD-WD mergers may vary between events, resulting in a wide distribution in both the properties of the accretion disk formed (which influences the prompt emission) and the rotation rate of the proto-magnetar (which determines the EE).  Event-to-event variability may also result from the viewing angle $\theta_{\rm obs}$ of the observer with respect to the rotation axis.  The spin-down power of the magnetar varies as $\dot{E} \propto \sin^{2}$(${\theta_{\rm obs}}$) for $\sigma_{\rm LC} \gg 1$; hence, the light curves in Figure \ref{fig:lc} remain reasonably accurate for moderately large $\theta_{\rm obs}$, but a viewer looking down the rotation axis ($\theta_{\rm obs} \sim 0^{o}$) may observe little or no EE.  Conversely, equatorial viewers would observe EE but no prompt spike; therefore, this model predicts a class of long duration X-ray flashes (XRFs) not associated with very massive star formation or accompanied by a SN.  Such an event may have already been observed.  GRB060428b is an XRF with a light curve similar to the EE from GRB060614 which was localized inside a red galaxy at redshift $z = 0.347$.  Assuming this galaxy is the host, GRB060428b released an isotropic energy $\sim 2\times 10^{50}$ ergs, comparable to EE of GRB060614, and showed no SN component at one month down to an optical brightness $\sim 20$ times fainter than SN1998bw (Perley et al., in prep).  

Because the EE flow is symmetric in azimuth, equatorial viewers should not observe a classic jet-break (although a more shallow break is possible; Thompson 2005).  Furthermore, although only $\Gamma \simgt 10$ material contributes to the EE in Figure \ref{fig:lc}, the magnetar releases $\sim 10^{52}$ ergs in earlier, mildly relativistic material ($\Gamma \sim 1-10$; see Fig.~\ref{fig:wind}) that is hoop-stress collimated along the PNS rotation axis.  This material may become visible as a radio transient as it slows down and becomes non-relativistic on a timescale of months to years.  If AIC or WD-WD mergers indeed produce SGRBEEs, these events should be bound to both early and late type galaxies; indeed, the well-known SGRBEEs 050709 and 050724 were localized to a star forming galaxy and an elliptical galaxy, respectively (Villasenor et al.~2005; Berger et al.~2005).  Unlike COMs, magnetar birth from AIC or WD-WD mergers should not produce strong gravitational wave emission, and because the magnetar will not collapse to a BH, its magnetic energy could power late-time X-ray flares.  

Although we have concentrated on spin-down-powered EE, isolated NS birth may produce the EE of SGRBEEs in other ways.  Specifically, the accretion disk produced by a WD-WD merger \emph{prior to collapse} (e.g., Yoon et al.~2007) may accrete onto the NS at later times, powering a bipolar outflow similar to that produced during the prompt accretion episode; in this case, the delay until EE reflects the accretion timescale at the WD radius ($\sim 10^{9}$ cm), which D07 estimate is $t_{\rm visc} \sim 100$ s for $\alpha \sim 0.1$.  Late-time accretion and spin-down powered EE can be distinguished based on the presence or absence, respectively, of a jet break and the observed ratio of SGRBEEs to off-axis, purely-EE XRFs.  Assuming that jets from the prompt and delayed accretion episodes are similarly collimated, SGRBEEs with accretion-powered EE should not be visible off-axis; by contrast, if the EE is powered by magnetar spin-down at least as many off-axis XRFs are expected as standard SGRBEEs. 

Finally, it is important to distinguish the observable signature of a magnetar produced by an AIC, WD-WD merger, or NS-NS merger from that produced by the core collapse of a massive star, which may instead produce a traditional long duration GRB (e.g., M07a).  The magnetic fields and rotation rates of the magnetars produced via these channels may differ, which would modify $\dot{E}$ and $\sigma$ of the wind (Fig.~\ref{fig:wind}) and hence its observable properties.  Although isolated magnetar spin-down may be comparatively simple because the proto-magnetar wind expands relatively freely into space, a free magnetar wind is nearly isotropic and so its emission is relatively weak and difficult to detect.  By contrast, the wind from a magnetar produced via core collapse is collimated into a bipolar jet by the overlying stellar mantle (Uzdensky $\&$ MacFadyen 2006; Komissarov $\&$ Barkov 2007; Bucciantini et al.~2007) and the observed emission can be much brighter due to the jet's modest opening solid angle.
\vspace{-0.75cm}
\section*{Acknowledgments}
We thank J.~Bloom, N.~Bucciantini, N.~Butler, D.~Perley, A.~Spitkovsky, and E.~Troja for helpful discussions and information.  EQ and BDM were supported by the David and Lucile Packard Foundation and a NASA GSRP Fellowship. 
\vspace{-0.75cm}

\label{lastpage}

\end{document}